\documentclass[12pt,oneside,letter]{article}
\usepackage[font=footnotesize,labelfont=bf]{caption}
\usepackage{tabularx}
\usepackage{color}
\usepackage{amsmath,amsfonts,amssymb,amscd}
\usepackage{graphicx}
\usepackage{cite}
\usepackage{booktabs}
\usepackage{paralist}
\usepackage{enumitem}
\usepackage[export]{adjustbox}
\usepackage{multicol}
\usepackage{gensymb}
\usepackage{steinmetz}
\usepackage{mathtools}
\usepackage{siunitx}
\usepackage[draft]{hyperref}
\usepackage{wrapfig}
\usepackage{subfig}

\usepackage[english]{babel}
\usepackage{blindtext}

\usepackage[compact]{titlesec}
\titlespacing{\section}{0pt}{*0}{*0}
\titlespacing{\subsection}{0pt}{*0}{*0}
\titlespacing{\subsubsection}{0pt}{*0}{*0}

\usepackage[left=1.0in,top=1.0in,right=1.0in,bottom=0.6in]{geometry}

\setdescription{leftmargin=\parindent,labelindent=\parindent}
\setdescription{leftmargin=0.5\parindent}

\usepackage{setspace}

\graphicspath{{./figures/}}

\usepackage{fancyhdr}
    \renewcommand{\headrulewidth}{0pt} 
    \renewcommand{\footrulewidth}{0pt}

\usepackage{lipsum}

\let\OLDthebibliography\thebibliography
\renewcommand\thebibliography[1]{
  \OLDthebibliography{#1}
  \setlength{\parskip}{0pt}
  \setlength{\itemsep}{0pt plus 0.3ex}
}

\usepackage{titlesec}

\titleformat{\section}
  {\normalfont\fontsize{12}{15}\bfseries}{\thesection}{1em}{}
\titleformat{\subsection}
  {\normalfont\fontsize{12}{15}\bfseries}{\thesubsection}{1em}{}

\begin{document}

\setlength{\belowdisplayskip}{0pt} \setlength{\belowdisplayshortskip}{0pt}
\setlength{\abovedisplayskip}{-10pt} \setlength{\abovedisplayshortskip}{-10pt}

\setlength{\footskip}{16pt}
\fancypagestyle{IEEEcopyright}{%
  \fancyhf{}%
  \renewcommand{\headrulewidth}{0pt}%
  \renewcommand{\footrulewidth}{0pt}%
  \fancyfoot[L]{\parbox[t]{\textwidth}{\scriptsize\linespread{1}\selectfont
    \copyright{} 2023 IEEE. Personal use of this material is permitted.
    Permission from IEEE must be obtained for all other uses, in any current or
    future media, including reprinting/republishing this material for advertising
    or promotional purposes, creating new collective works, for resale or
    redistribution to servers or lists, or reuse of any copyrighted component of
    this work in other works.\\
    DOI: 10.1109/ECCE53617.2023.10362032}}%
}

\title {\large  \bf \vspace{-8ex}
Stability of Droop-Controlled Low-Frequency Transmission Lines
\vspace{-4ex}}

\author{%
\setstretch{1.0}\hspace*{-\tabcolsep}%
\begin{minipage}[t]{0.315\textwidth}\centering
\normalsize Rafael Castillo-Sierra\\[1pt]
{\scriptsize\itshape Dept. of Electrical and Computer Eng.}\\
{\scriptsize\itshape University of Wisconsin-Madison}\\
{\scriptsize Madison, USA}\\
{\scriptsize castillosier@wisc.edu}
\end{minipage}\hspace{0.009\textwidth}%
\begin{minipage}[t]{0.315\textwidth}\centering
\normalsize Giri Venkataramanan\\[1pt]
{\scriptsize\itshape Dept. of Electrical and Computer Eng.}\\
{\scriptsize\itshape University of Wisconsin-Madison}\\
{\scriptsize Madison, USA}\\
{\scriptsize giri@engr.wisc.edu}
\end{minipage}\hspace{0.009\textwidth}%
\begin{minipage}[t]{0.351\textwidth}\centering
\normalsize Dionisio Ramirez\\[1pt]
{\scriptsize\itshape Escuela Superior de Ingenieros Industriales}\\
{\scriptsize\itshape Universidad Polit\'ecnica de Madrid}\\
{\scriptsize Madrid, Spain}\\
{\scriptsize dionisio.ramirez@upm.es}
\end{minipage}\hspace*{-\tabcolsep}%
\vspace{-1ex}}

\date{}
\maketitle
\thispagestyle{IEEEcopyright}
\pagestyle{empty}

\renewcommand{\figurename}{Fig.}

\vspace{-1em}
\begin{spacing}{1.2}
\textbf{Abstract—} Low-Frequency AC (LFAC) transmission systems employing power converters are being considered for a varity of applications. This work studies the small-signal stability of an LFAC transmission line controlled by the Droop Control Strategy. Eigenvalue Analysis is used to determine how the controller droop gains, operating frequency, transmission line parameters, and the operating point affect system stability. The results show that the overall system's dynamic is governed by the sum of the droop gains of the AC/AC converters. Analytical results that give insights on how the different system's parameters affect the critical stability point are presented. The results indicate that system stability is affected by the line's length, the line's R/X ratio, operating frequency and voltage.
\end{spacing}
\vspace{0.1em}

\section{Introduction}\label{sec:Intro}

\begin{wrapfigure}{l}{0.40\textwidth}
\vspace{-5pt}
\centering
\includegraphics[width=0.40\textwidth]{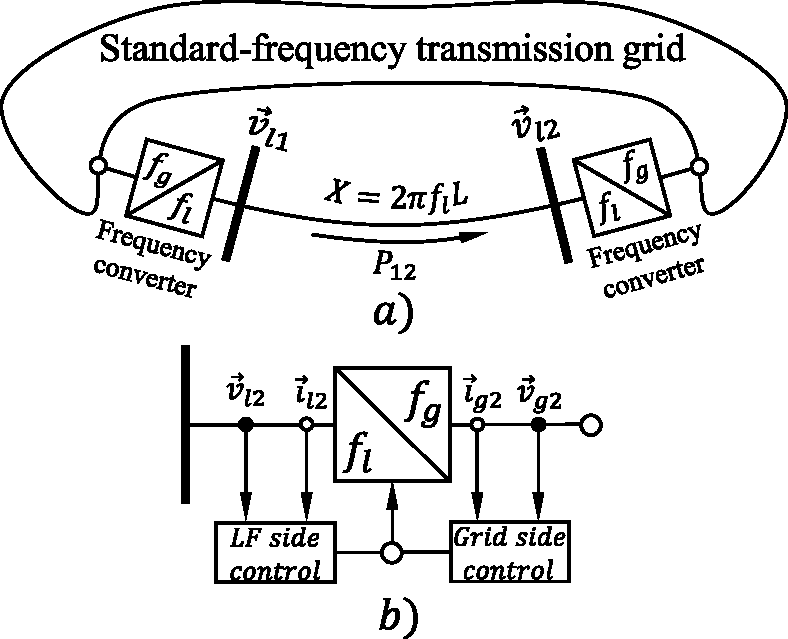}
\caption{a) LFAC transmission line embedded into a standard-frequency transmission system. b) Converter's controls structure.}
\vspace{-10pt}
\label{Fig01}
\end{wrapfigure}

Low-Frequency AC Transmission (LFAC) refers to transport power at a frequency lower than the standard 50/60Hz. This technology is an alternative for long-distance transmission and offshore wind farm interconnection since the reduction of the transmission frequency mitigates operational constraints such as voltage drops \cite{7741422}, and transient stability \cite{6863297}. Consequently, transmission corridors can be used up to the conductor's thermal limit. The practical implementation of an LFAC line in a transmission system is outlined in Fig. \ref{Fig01}a. The conceptual idea is to reduce the frequency of individual transmission lines or sections of the transmission system by using AC/AC converters that serve as the interface between the two systems. These converters are responsible for forming of the low-frequency network through the implementation of control schemes such as Following-Follower Control \cite{9292825}, Droop Control \cite{9224834} or Virtual Synchronous Machine Strategy \cite{9056566}, similar to microgrids.

Very few studies have focused on studying the stability of power converter control in LFAC lines. In addition, only the Following-Follower Strategy has been used in these studies \cite{RUDDY2018220,9697599}. Therefore, this work aims to present a small-signal stability analysis of a Droop-Controlled LFAC system. With this analysis, it is possible to identify how the line's parameters, the operating point variables, and the power-frequency droop gains of converters at the two ends of a point-point line affect the system's stability.

\section{Control Architecture Of The LFAC Branch}


Each frequency converter of the LFAC line has two controllers, as shown in Fig. \ref{Fig01}b.



The Grid-Side controller keeps the converter's stored energy stable typically, by cascading two loops. The inner loop controls the \textit{dq}-axes currents based on the power commands. The outer loop regulates the converter's DC voltage by modifying the \textit{d}-axis current command.



The LF-Side controller has three loops to form the LFAC network at the desired voltage and frequency. The inner loops control the \textit{dq}-axes currents and the terminal voltage, while the outer loop controls the power flow quantities, which is the subject of this paper.

The voltage and frequency regulation is carried out through the outer power loop, which is realized by the Droop Control Strategy. This control scheme allows the converters to share the responsibility for regulating frequency and voltage. In this sense, as seen in (\ref{Eq01}), the frequency is regulated to its commanded value $(\omega^*)$ through the active power error $(P^* - P)$ and the power-frequency droop gain $m_p$ (in $rad/s/W$). The voltages are regulated through the reactive power error $(Q^* - Q)$ and the reactive power-voltage droop gain $m_q$ (in $V/VAR$).

\begin{equation}
\omega=\omega^\ast+m_p\left(P^\ast-P\right),\ \ V=V^\ast+m_q\left(Q^\ast-Q\right)
\label{Eq01}
\end{equation}

\section{Dynamic Modeling and Linearization of The LFAC System}

This section presents the LFAC line modeling under the following assumptions. ($i$) The Grid-Side controller keeps the converters' stored energy stable, implying stiff inner DC voltages in both converters. ($ii$) The current and voltage loops in the LF-Side controller are fast enough, so they can be omitted. ($iii$) The transmission line's dynamic properties will be considered. ($iv$) Each converter's reactive power droop gains $m_q$ is set zero. ($v$) The measurement filters are not considered. These aspects will be considered in the future.

\begin{wrapfigure}{l}{0.60\textwidth}
\vspace{0pt}
\centering
\includegraphics[width=0.60\textwidth]{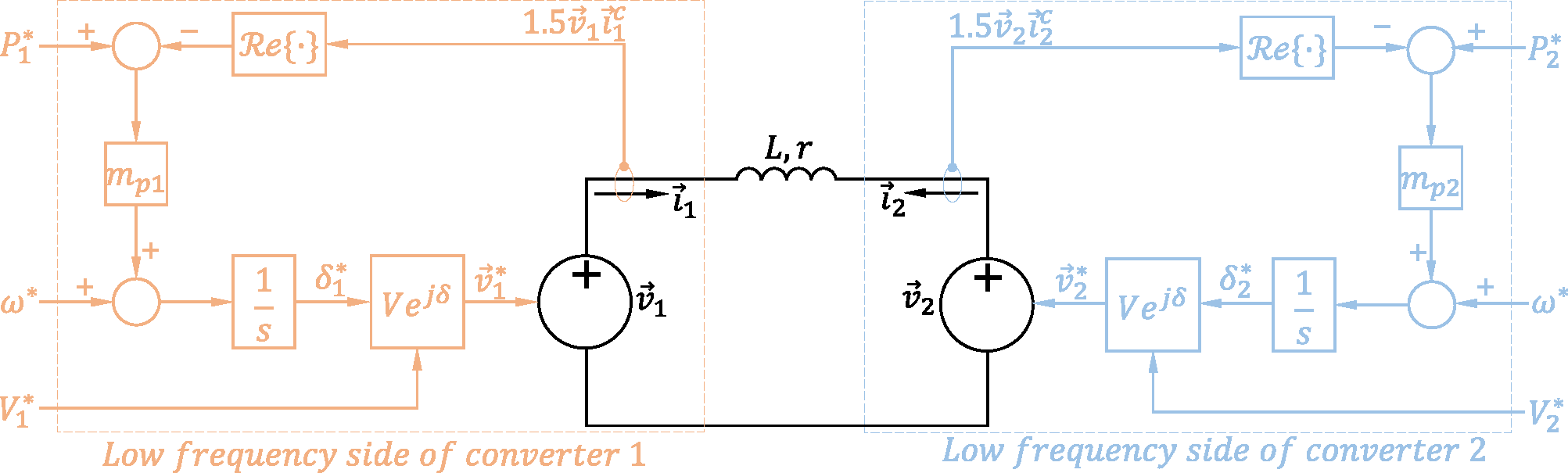}
\caption{Reduced-order complex-variable circuit of the low-frequency side with the power-frequency droop loops highlighted in orange and blue.}
\vspace{-20pt}
\label{Fig03}
\end{wrapfigure}

On the basis of these assumptions, the transmission line's low-frequency side model is shown in Fig. \ref{Fig03}. The transmission line is represented by an $r$-$L$ branch, and the sources $\vec{v} _1$ and $\vec{v} _2$ represent the converter voltages at the LF-side terminals. The loops highlighted in orange and blue correspond to the power-frequency droop loops represented by the expressions in (\ref{Eq02}). $m_{p1}$ and $m_ {p2}$ are the power-frequency droop gains of each converter.

\begin{equation}
\omega_1=\omega^\ast+m_{p1}\left(P_1^\ast-P_1\right),\ \ \omega_2=\omega^\ast+m_{p2}\left(P_2^\ast-P_2\right)
\label{Eq02}
\end{equation}

The voltage angle ($\delta_1$ and $\delta_2$) comes from the integration of the frequencies shown in (\ref{Eq02}). The voltage magnitudes ($V_1$ and $V_2$) are maintained constant at the commanded values ($V^*_1$ and $V^*_2$) since the reactive power-voltage droop gains ($m_{q1}$ and $m_{q2}$) have been set to zero.


The low-frequency side dynamics comprises the transmission line's dynamics shown in (\ref{Eq03}) and the rate of change of the transmission angle shown in (\ref{Eq04}). The rotating reference frame speed ($\omega_r$) is tied to the voltage vector $\vec{v}_1$ speed ($\omega_1$).

\begin{equation}
v_{1d}=v_{2d}+L\frac{d}{dt}i_{1d}-\omega_rLi_{1q}+ri_{1d},\ \ v_{1q}=v_{2q}+L\frac{d}{dt}i_{1q}+\omega_rLi_{1d}+ri_{1q}
\label{Eq03}
\end{equation}

\begin{equation}
\frac{d}{dt}\delta_{21}=\omega_2-\omega_1=m_{p2}\left(P_2^\ast-P_2\right)-m_{p1}\left(P_1^\ast-P_1\right)
\label{Eq04}
\end{equation}


The linearized version of (\ref{Eq03}) and (\ref{Eq04}) is presented in (\ref{Eq05}) in state-space representation. Variables with subscript "0" represent voltages, currents, or powers evaluated at the operation point at which the linearization was made. Specifically, $P_ {10}$ and $Q_ {10}$ are the powers measured at the converter 1 terminals, while $Q_ {20}$ is the reactive power in converter 2.

\begin{equation}
\resizebox{0.93\linewidth}{!}{$\displaystyle
\frac{d}{dt}\left[\begin{matrix}{\widetilde{i}}_{1d}\\{\widetilde{i}}_{1q}\\{\widetilde{\delta}}_{21}\\\end{matrix}\right]=\left[\begin{matrix}-\frac{r}{L}+m_{p1}Q_{10}&\omega_0&\frac{V_{2q0}}{L}\\-\omega_0+m_{p1}P_{10}&-\frac{r}{L}&-\frac{V_{2d0}}{L}\\\frac{3}{2}\left(m_{p1}V_{10}+m_{p2}V_{2d0}\right)&\frac{3}{2}m_{p2}V_{2q0}&m_{p2}Q_{20}\\\end{matrix}\right]\left[\begin{matrix}{\widetilde{i}}_{1d}\\{\widetilde{i}}_{1q}\\{\widetilde{\delta}}_{21}\\\end{matrix}\right]+\left[\begin{matrix}m_{p1}I_{1q0}&0&I_{1q0}\\-m_{p1}I_{1d0}&0&-I_{1d0}\\-m_{p1}&m_{p2}&0\\\end{matrix}\right]\left[\begin{matrix}{\widetilde{P}}_1^\ast\\{\widetilde{P}}_2^\ast\\{\widetilde{\omega}}^\ast\\\end{matrix}\right]
$}
\label{Eq05}
\end{equation}

\section{Effect Of Power-Frequency Droop Gains $m_{p1}$ and $m_{p2}$ On System’s Stability}

This section analyzes how the interaction between the converters' power-frequency droop loops and the transmission line dynamics affects the stability. The system's stability is evaluated through the Eigenvalue Analysis of the characteristic equation shown in (\ref{Eq06}), where $s$ is the Laplace operator, $I$ is the identity matrix, and $A$ is the state matrix in (\ref{Eq05}).

\begin{equation}
\left|sI-A\right|=s^3+\left(2\frac{r}{L}\right)s^2+\left(\frac{r^2}{L^2}+\omega_0^2\right)s+\frac{3}{2}\frac{\omega_0}{L}\left(m_{p1}V_{10}^2+m_{p2}V_{20}^2\right)
\label{Eq06}
\end{equation}



\begin{wrapfigure}{l}{0.40\textwidth}
\vspace{0.0pt}
\centering
\includegraphics[width=0.40\textwidth]{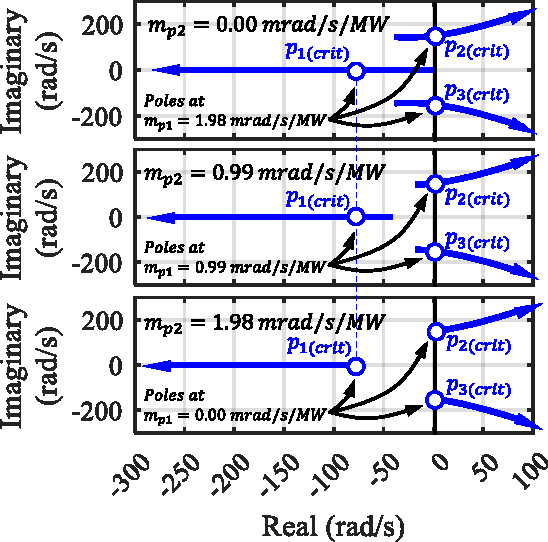}
\caption{Pole migration by varying $m_{p1}$ for three different values of $m_{p2}$.}
\vspace{-20pt}
\label{Fig04}
\end{wrapfigure}

The effect of the droop gains in stability is graphically presented in Fig. \ref{Fig04}. This plot shows the pole migration by varying $m_ {p1}$ from $0.0$ to $20mrad/s/MW$. Also, three versions of the pole migration are plotted for three values of $m_{p2}$, $0.00$, $0.99$, and $1.98mrad/s/MW$.

As expected, there are three roots, one is always real and negative regardless of the values of $m_{p1}$ and $m_ {p2}$. The remaining two are complex conjugate roots that gradually migrate to the right side of the complex plane as $m_ {p1}$ grows. By analyzing the three plots, it is observed that the marginally stable operating point (white dots) always happens when the sum $m_{p1}+m_{p2}$ equals a fixed value of $1.98mrad/s/MW$ for the case study. This indicates that the system's dynamics is governed by the sum of the two droop gains, which is coined in this study as the \textit{Global Droop Gain} and defined as shown in (\ref{Eq07}).

\begin{equation}
m_{pg}=m_{p1}+m_{p2}
\label{Eq07}
\end{equation}
The reason why (\ref{Eq07}) governs the system's dynamics can be explained by analyzing the independent coefficient of (\ref{Eq06}). by assuming that $V_ {10}$ and $V_{20}$ are approximately equal (given the voltage regulation limits in transmission systems), the independent coefficient will have the form $\frac{3}{2}\frac{\omega_0}{L}\left(m_{p1}+m_{p2}\right)V_0^2$ where the presence of the Global Droop Gain is clear.

\section{Critical Global Droop Gain}

Fig. \ref{Fig04} also shows that the marginally stable operation point happens when $m_{pg}$ is equal to a fixed amount regardless of how it is subdivided between $m_{p1}$ and $m_{p2}$. From the characteristic equation evaluated at the marginally stable point, it is possible to obtain the symbolic expression presented in (\ref{Eq08}) of the so-called \textit{Critical Global Droop Gain}, which is defined as the global gain $m_{pg}$ that makes the system marginally stable.

\begin{equation}
m_{pg(critical)}=\frac{4L}{3}\left(\frac{\omega_l}{\omega_0}\right)\frac{\omega_l^2+\omega_0^2}{k_2V_{20}^2+k_1V_{10}^2},\ \ \omega_l=\frac{r}{L},\ \ k_1=\frac{m_{p1}}{m_{pg}},\ \ \ k_2=\frac{m_{p2}}{m_{pg}}
\label{Eq08}
\end{equation}

It is observed in (\ref{Eq08}) that $m_{pg(critical)}$ depends exclusively on the line parameters, the operating frequency and voltage. This expression also provides information on how low-frequency line parameters affect the critical stability limit. Here are some observations:

\textbf{1)} $m_{pg(critical)}$ is proportional to inductance $L$ which is equivalent to being proportional to the line's length. This implies that short lines have lower stability region than long ones.

\textbf{2)} The ratio $\frac{\omega_l}{\omega_0}$ is a measure of the line's $R/X$ ratio. This indicates that inductive transmission lines have lower values of $m_{pg(critical)}$ than those lines with more resistive components.

\textbf{3)} There is a directly proportional relationship between $m_{pg(critical)}$ and $\omega_0$ if $\omega_0>>\omega_l$.

\textbf{4)} The inverse of the square voltage affects the critical gain $m_{pg(critical)}$. This means that transmitting at higher voltages has a negative effect on the system's stability range.

\section{Conclusions and Future Work}

This work presents the small-signal stability analysis of a low-frequency AC transmission line controlled by Droop Control. Using the Eigenvalue Analysis, it was possible to identify that the LFAC line dynamic is governed by the sum of the converters' droop gains. Also, a symbolic expression was obtained that predicts the Global Droop Gain that makes the system marginally stable. This expression shows that the stability limit is a function of the line's length, the line's R/X ratio, the operating frequency, and voltage. This dependency shows that the system itself imposes the constraints regarding how much droop gain is allowed in the converters' controllers. The final paper will provide more details on system modeling and mathematical derivation of the global critical gain expression. In addition, the stability analysis will be expanded by including the effect of reactive power-voltage loops. Finally, simulation and experimental results will be provided to verify the findings in this digest.

{\setstretch{1}\vspace{\baselineskip}\small

}

\end{document}